\documentclass[12pt]{article}
\title{Tensor susceptibility calculated in the hadronization process}
\author{Hongting Yang \\
       {\small\sl Department of Physical Science and Technology, College of Science}\\
       {\small\sl Wuhan University of Technology, Wuhan 430070, P.R. China}}
\date{}
\usepackage{graphicx}
\begin{document}
\maketitle
\begin{abstract}
The tensor susceptibility of QCD vacuum is calculated in the global color symmetry model. The input
parameters for gluon propagators are determined via the simplified equation for calculating the
pion decay constant. The reason for the great discrepancy between our results and those from QCD
sum rules and from chiral constituent quark model is discussed.
\end{abstract}
{\it E-mail}\/: yht@mail.whut.edu.cn
\newpage

The tensor susceptibility of QCD vacuum, like the quark condensate or the gluon condensate,
reflects the non-perturbative aspects of the QCD vacuum directly. It is argued that, the tensor
susceptibility is related to a chiral-odd spin-dependent structure function that can be measured in
the polarized Drell-Yan process~(\cite{HX95}-\cite{JD79}). The earlier estimations for the value of
tensor susceptibility were obtained by QCD sum rules techniques~(\cite{HX96}-\cite{AS00}) or from
chiral constituent quark model~\cite{WM98}. Two decades calculations show that, the global color
symmetry model (GCM)~\cite{RC85} describes the nonperturbative aspects of strong interaction
physics and hadronic phenomena at low energies quite well~(\cite{CR88}-\cite{HJ03}), we naturally
expect that GCM is applicable in the estimation of the tensor susceptibility of QCD vacuum. Recent
investigation shows that, the value of tensor susceptibility calculated from an effective
quark-quark interaction is much smaller than the others~\cite{HH03}. Questions are therefore
arising. What is the reason for this discrepancy? Is anything wrong with GCM? This letter aims to
answer these questions.

The QCD partition function for massless quarks in Euclidean space can be written as
\begin{equation}
   {\cal Z}=\int{\cal D}\bar{q}{\cal D}q{\cal D}Ae^{-S[\bar{q},q,A]}
\end{equation}
with the QCD action
\begin{equation}
   S[\bar{q},q,A]=\int{\rm d}x[\bar{q}(x)({\not\!\partial}-ig{\not\!\!A})q(x)+
   {1\over 4}F^a_{\mu\nu}F^a_{\mu\nu}],
\end{equation}
where $A_{\mu}=A_{\mu}^a{\lambda^a\over 2}$,
$F^a_{\mu\nu}=\partial_\mu{A^a_\nu}-\partial_\nu{A^a_\mu}+gf^{abc}{A^b_\mu}{A^c_\nu}$. By
introducing the functional $W[J]$ defined as
\begin{equation}
  e^{W[J]}\equiv\int{\cal D}A\,\exp\left(\int{\rm d}x(-{1\over 4}F^a_{\mu\nu}F^a_{\mu\nu}+
  J^a_{\mu}A^a_{\mu})\right),
\end{equation}
the QCD partition function can be rewritten as
\begin{equation}
  {\cal Z}=\int{\cal D}\bar{q}{\cal D}q\,e^{-\int{\rm d}x\bar{q}(x){\not\,\partial}q(x)}
  e^{W[ig\bar{q}\gamma_\mu{\lambda^a\over2}q]}.
\end{equation}
The functional $W[J]$ has the expansion
\begin{equation}
    W[J]={1\over 2}\int{\rm d}x{\rm d}yJ^a_{\mu}(x)D^{ab}_{\mu\nu}(x,y)J^b_{\nu}(y)+W_R[J],
\end{equation}
where $D^{ab}_{\mu\nu}(x,y)=D^{ab}_{\mu\nu}(x-y)$ is the gluon 2-point Green's function, and
$W_R[J]$ involves the higher order $n(\ge3)$-point Green's functions. The GCM is obtained through
the truncation of the functional $W[J]$ in which only $D^{ab}_{\mu\nu}(x,y)$ is retained. This
model maintains global color symmetry of QCD, but the local color SU(3) gauge invariance is lost.
For simplicity we use a Feynman-like gauge $D^{ab}_{\mu\nu}(x-y)=\delta_{\mu\nu}\delta^{ab}D(x-y)$.
The important dynamical characteristics of local color symmetry are included in $D(x)$.  The exact
form of $D(x)$ is not well known. Instead, we use a phenomenological gluon propagator, which is
required to exhibit the properties of asymptotic freedom and infrared slavery. The justification of
such truncation relies on the successes of various calculations.

The partition function of this truncation can be given as~\cite{RC85,T97}
\begin{equation}     \label{pfGCM}
   {\cal Z}_{\rm GCM}=\int{\cal D}\bar{q}{\cal D}q\exp\left(-\int{\rm d}x\bar{q}{\not\!\partial}q-
   \frac{g^2}{2}\int{\rm d}x{\rm d}yj^a_{\mu}(x)D^{ab}_{\mu\nu}(x-y)j^b_{\nu}(y)\right ),
\end{equation}
with the quark color current $j^a_{\mu}(x)=\bar{q}(x)\gamma_{\mu}\frac{\lambda^a}{2}q(x)$, or
equivalently,
\begin{equation}
   {\cal Z}_{\rm GCM}=\int{\cal D}\bar{q}{\cal D}q{\cal D}Ae^{-S_{\rm GCM}[\bar{q},q,A]}
\end{equation}
with the GCM action
\begin{equation}
   S_{\rm GCM}[\bar{q},q,A]=\int{\rm d}x[\bar{q}(x)({\not\!\partial}-ig{\not\!\!A})q(x)+
   \int{\rm d}x{\rm d}y{1\over 2}A_{\mu}^a(x)[D^{ab}_{\mu\nu}(x-y)]^{-1}A_{\nu}^b(y).
\end{equation}
By the standard bosonization procedure, the resulting expression for the partition function in
terms of the bilocal field integration is
\begin{equation}
  {\cal Z}_{\rm GCM}=\int{\cal D}{\cal B}^{\theta}\exp\left(-S[{\cal B}^{\theta}]\right),
\end{equation}
where the action is given by
\begin{equation}
   S[{\cal B}^{\theta}]=-{\rm Tr\,Ln}[G^{-1}]+\int{\rm d}x{\rm d}y
   \frac{{\cal B}^{\theta}(x,y){\cal B}^{\theta}(y,x)}{2g^2D(x-y)}\,,
\end{equation}
and the quark inverse Green's function $G^{-1}$ is defined as
\begin{equation}
   G^{-1}(x,y)={\not\!\partial}\delta(x-y)+\Lambda^{\theta}{\cal B}^{\theta}(x,y)\,.
\end{equation}
Here the quantity $\Lambda^{\theta}$ arises from Fierz reordering of the current-current
interaction term in Eq.~(\ref{pfGCM})
\begin{equation}
 \Lambda^{\theta}_{ji}\Lambda^{\theta}_{lk}=(\gamma_{\mu}\frac{\lambda^a}{2})_{jk}
 (\gamma_{\mu}\frac{\lambda^a}{2})_{li}
\end{equation}
and is the direct product of Dirac, flavor SU(2) and color matrices:
\begin{equation}
  \Lambda^{\theta}={1\over 2}(I_D,i\gamma_5, \frac{i}{\sqrt{2}}\gamma_{\mu},
  \frac{i}{\sqrt{2}}\gamma_{\mu}\gamma_5)\otimes({1\over\sqrt{2}}I_F,
  {1\over\sqrt{2}}{\vec{\tau}_F})\otimes({4\over 3}I_c, \frac{i}{\sqrt{3}}\lambda^a_c)\,.
\end{equation}
Here we consider $N_F=2$ flavors as in Ref.~\cite{WM98}.

The vacuum configurations are defined by minimizing the bilocal action: $\left. \frac{\delta
S[{\cal B}]}{\delta {\cal B}} \right |_{{\cal B}_0}=0,$ which gives
\begin{equation}
  {\cal B}^{\theta}_0(x)=g^2D(x){\rm tr}[G_0(x)\Lambda^{\theta}]\,.
\end{equation}
These configurations provide self-energy dressing of the quarks through the definition
$\Sigma(p)\equiv\Lambda^{\theta}{\cal B}^{\theta}_0(p) =i{\not\!p}[A(p^2)-1]+B(p^2)$. According to
Ref.~\cite{RC85}, the self-energy functions $A$ and $B$ satisfy the Dyson-Schwinger equations,
\begin{eqnarray}
   [A(p^2)-1]p^2 &=&{8\over 3}\int\frac{{\rm d}^4q}{(2\pi)^4}\ g^2D\left((p-q)^2\right)\
   \frac{A(q^2)q\cdot p}{q^2A^2(q^2)+B^2(q^2)}\,,  \label{DSE1}  \\ \label{DSE2}
   B(p^2) &=&{16\over 3}\int\frac{{\rm d}^4q}{(2\pi)^4}\ g^2D\left((p-q)^2\right)\
   \frac{B(q^2)}{q^2A^2(q^2)+B^2(q^2)}\,.
\end{eqnarray}
The quark Green's function at ${\cal B}^{\theta}_0$ is given by
\begin{equation}
  G_0(x,y)=G_0(x-y)=\int\frac{{\rm d}^4p}{(2\pi)^4}
  \frac{-i{\not\!p}A(p^2)+B(p^2)}{p^2A^2(p^2)+B^2(p^2)}e^{ip\cdot(x-y)}\,.
\end{equation}

The hadron properties follow from considering deviations from these vacuum configurations. If we
consider only the isoscalar $\sigma(x)$ and isovector $\vec{\pi}(x)$ fields, the approximate
local-field effective action can be taken as~\cite{RC85}
\begin{eqnarray}
   S[\sigma,\vec{\pi}] &=& -{\rm Tr\,Ln}\left\{{\not\!\partial}A(x-y)+m\delta^{(4)}(x-y)+
      V[\sigma,\vec{\pi}]B(x-y)\right\} \nonumber\\
  & &{}+{1\over2}\int{\rm d}^4z[\sigma^2(z)+\vec{\pi}^2(z)]\int{\rm d}^4wB(w){\rm tr}[G(w)]\,,
\end{eqnarray}
where $V[\sigma,\vec{\pi}]=\sigma({x+y\over2})+ i\gamma_5\vec{\pi}({x+y\over2})\cdot\vec{\tau}$ and
$m$ is the quark bare mass. Expanding the spectrum of $S[\sigma,\vec{\pi}]$ to second order about
its minimum $S[1,0]$ with $\sigma(x)=1+\delta(x)$
\begin{equation}
   S[1+\delta(x),\vec{\pi}(x)]-S[1,0]={1\over2}{f_\delta}^2\int\left[(\partial_\mu\delta)^2+
   {m_\delta}^2\delta^2\right]{\rm d}^4z+{1\over2}{f_\pi}^2\int\left[(\partial_\mu\vec{\pi})^2+
   {m_\pi}^2\vec{\pi}^2\right]{\rm d}^4z+\cdots\,,
\end{equation}
it is found that
\begin{equation}
   {m_\delta}^2=\frac{3}{2\pi^2{f_\delta}^2}\int_0^\infty s{\rm d}s\ \frac{B^2(s)[B^2(s)-sA^2(s)]}
   {[sA^2(s)+B^2(s)]^2}\,,
\end{equation}
\begin{equation}  \label{fdelta}
   {f_\delta}^2=\frac{3}{8\pi^2}\int_0^\infty s{\rm d}s\ \frac{A^2(s)B^2(s)}{[sA^2(s)+B^2(s)]^3}
   \left\{2sA^2(s)+\frac{B^2(s)[sA^2(s)-B^2(s)]}{sA^2(s)+B^2(s)}\right\}\,;
\end{equation}
and
\begin{equation}
   {m_\pi}^2=\frac{3m}{2\pi^2{f_\pi}^2}\int_0^\infty s{\rm d}s\ \frac{B(s)}{sA^2(s)+B^2(s)}\,,
\end{equation}
\begin{equation}  \label{fpi1}
   {f_\pi}^2=\frac{3}{8\pi^2}\int_0^\infty s{\rm d}s\ \frac{A^2(s)B^2(s)}{[sA^2(s)+B^2(s)]^2}
   \left[2+\frac{B^2(s)}{sA^2(s)+B^2(s)}\right]\,.
\end{equation}
In Eqs.~(\ref{fdelta}) and (\ref{fpi1}), all those terms involving the derivatives of $A(s)$ and
$B(s)$ with respect to $s$ are neglected. Let $A(s)=1$, $B(s)=M$, with $M$ the mass of the
constituent quark, Eq.~(\ref{fpi1}) reduces to
\begin{equation}  \label{fpi2}
   {f_\pi}^2=\frac{3}{8\pi^2}\int_0^\infty s{\rm d}s\ \frac{M^2}{(s+M^2)^2}
   \left(2+\frac{M^2}{s+M^2}\right)\,.
\end{equation}
When $s$ approaches infinity, the integrand of Eq.~(\ref{fpi2}) behaves like $2M^2/s$, which
reproduce the result of Ref.~\cite{WM98} strictly. While ${s\rightarrow 0}$, it behaves like
$3s/M^2$ rather than $2s/M^2$ of Ref.~\cite{WM98}. This difference arises from the second term in
the brackets of Eq.~(\ref{fpi2}). Can this difference bring some serious problems? Let put this
question aside at present.

The tensor susceptibility $\chi$ is defined as~\cite{HX96}
\begin{equation}
  \chi\equiv\frac{\Pi_{\chi}(0)}{6\langle\bar{q}q\rangle}\,,
\end{equation}
where
\begin{equation} \label{qqbar}
  \langle\bar{q}q\rangle=-\frac{3}{4\pi^2}\int^\infty_0s{\rm d}s\ \frac{B(s)}{sA^2(s)+B^2(s)}
\end{equation}
is the quark condensate, we need only to calculate \(\Pi_{\chi}(0)\) defined as~\cite{HH03}:
\begin{equation}  \label{chi}
  {1\over 12}\Pi_{\chi}(0)\equiv-\frac{3}{4\pi^2}
   \int^\infty_0s{\rm d}s\left [\frac{B(s)}{sA^2(s)+B^2(s)}\right ]^2.
\end{equation}
In this equation, if we let $A(s)=1$, $B(s)=M$, the result of Ref.~\cite{WM98} is strictly
reproduced.

Before the numerical calculation of tensor susceptibility, we make an analysis of the nontrivial
solutions $A(s)$ and $B(s)$ to the Dyson-Schwinger equations (\ref{DSE1}) and (\ref{DSE2}). Because
the phenomenological gluon propagators exhibit the properties of asymptotic freedom and infrared
slavery, when $s=p^2$ increases from 0 to $\infty$, $B(s)$ ($\ge0$) decreases from some finite
non-zero value to zero, while $A(s)$ decreases from some non-zero value down to 1. Therefore,
$A(s)$ is universally greater than 1. The replacement of $A(s)$ by 1 in Eq.~(\ref{chi}) will lead
an increase in the value of $\Pi_{\chi}(0)/12$, or in other words, the value calculated from
Eq.~(\ref{chi}) is expected to be smaller than that obtained by the others. This is verified in the
recent calculation~\cite{HH03}, and will be further checked in the calculation below.

The tactics in calculation is similar to that of Ref.~\cite{HH03}. The input parameters are
adjusted to reproduce the pion decay constant in the chiral limit $f_\pi=87$~MeV via
Eq.~(\ref{fpi1}).

\begin{figure}[hb]
\begin{center}
\rotatebox{0}{\includegraphics[height=6.5cm]{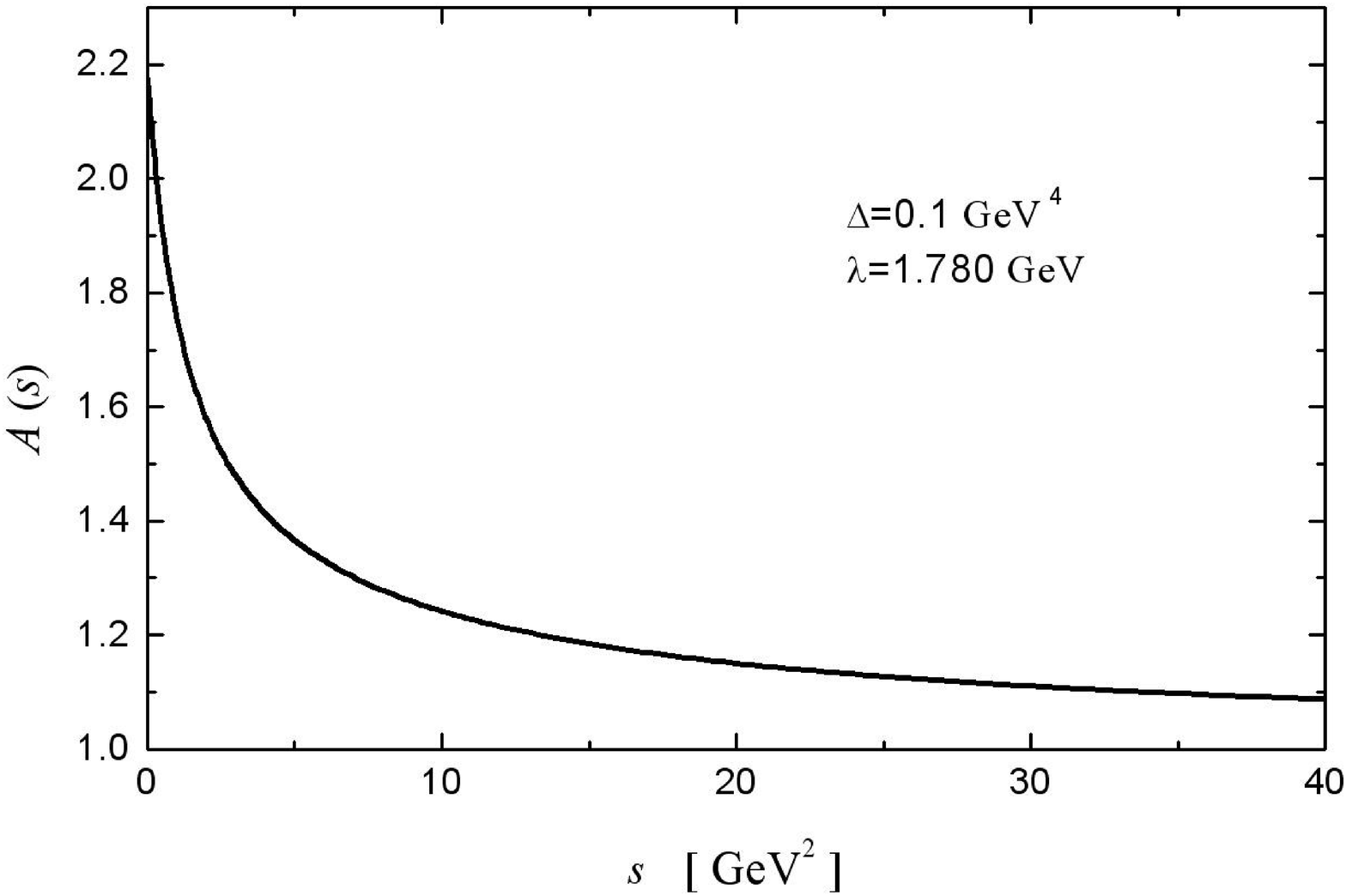}}\caption{\label{fig.as} The self-energy
function $A(s)$ as a function of $s$ for the gluon propagator
$g^2D(s)=\frac{48\pi^2}{29}\frac{\lambda^2}{s^2+\Delta}$, with $\Delta=0.1\,{\rm GeV}^4$,
$\lambda=1.780\,{\rm GeV}$}
\includegraphics[height=6.5cm]{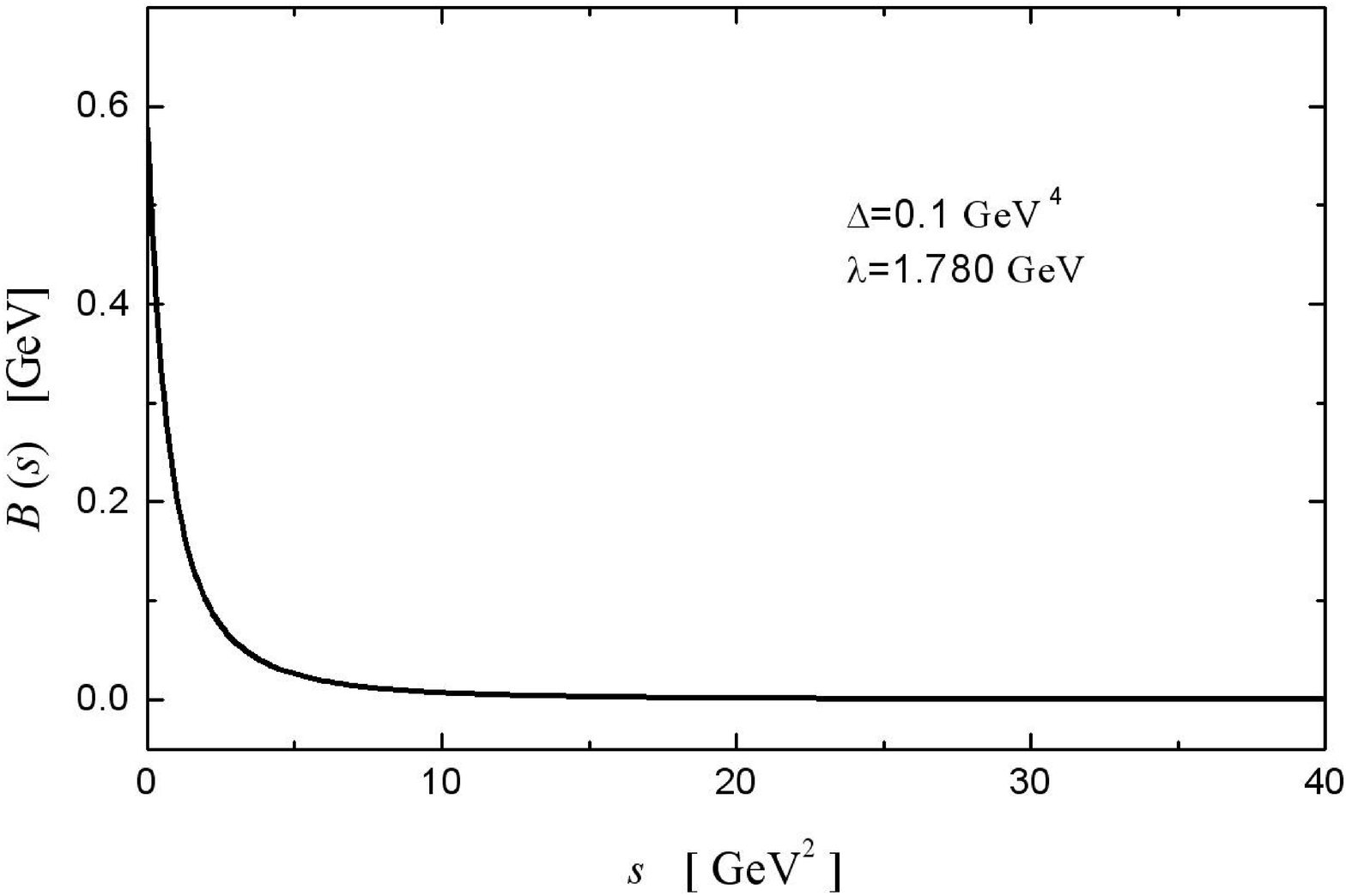}\caption{\label{fig.bs}The self-energy
function $B(s)$ as a function of $s$ for the gluon propagator
$g^2D(s)=\frac{48\pi^2}{29}\frac{\lambda^2}{s^2+\Delta}$, with $\Delta=0.1\,{\rm GeV}^4$,
$\lambda=1.780\,{\rm GeV}$}
\end{center}
\end{figure}

To give a convincible conclusion, we choose three different gluon propagators. The ultraviolet
behavior of these model gluon propagators are different from that in QCD~\cite{PC97,KH99}. They are
model 1:
\begin{equation}
   g^2D(s)=4\pi^2d\frac{\lambda^2}{s^2+\Delta}\,,
\end{equation}
 model 2:
\begin{equation}
g^2D(s)=3\pi^2\frac{\lambda^2}{\Delta^2}e^{-{s\over\Delta}} +\frac{4\pi^2d}{s\ln[s/\Lambda^2+e]}\,,
\end{equation}
and model 3:
\begin{equation}
g^2D(s)=4\pi^2d\,\frac{\lambda^2}{s^2+\Delta} +\frac{4\pi^2d}{s\ln[s/\Lambda^2+e]}\,.
\end{equation}
Here $d=12/(33-2N_f)=12/29$ and $\Lambda=200$ MeV. For model 1, the self-energy functions $A(s)$
and $B(s)$ varying with $s$ are showed respectively in Figs.~\ref{fig.as} and \ref{fig.bs}, with
the input parameters $\Delta=0.1\,{\rm GeV}^4$, $\lambda=1.780\,{\rm GeV}$. Obviously, $A(s)$ is
not less than 1. In Table~\ref{Tablechi} the values of $\Pi_{\chi}(0)/12$ for model 1 are
displayed, and the corresponding values for quark condensate $\langle\bar{q}q\rangle$ are also
listed. It should be noted that the values for quark condensate are roughly around those obtained
from QCD sum rules (see Ref.~\cite{T97} and references therein), due to the term
$B(s)/(sA^2(s)+B^2(s))$ in Eq.~(\ref{qqbar}) against the term $[B(s)/(sA^2(s)+B^2(s))]^2$ in
Eq.~(\ref{chi}) .
\begin{table}[ht]
\caption{\label{Tablechi} The values of $\Pi_{\chi}(0)/12$ for model 1 with Eq.~(\ref{fpi1}) used
to calculate $f_\pi$. The quark condensate $\langle\bar{q}q\rangle$ is also presented.}
\begin{center}
\begin{tabular}{cccc}
\hline $\Delta\:[{\rm GeV}^4]$ & $\lambda\:[{\rm GeV}]$ & $-\langle\bar{q}q\rangle^{1/3}\:[{\rm
MeV}]$ & $\Pi_{\chi}(0)/12\: [{\rm GeV}^2]$ \\ \hline
 $10^{-1}$       &1.780 & 279   &  -0.0017         \\
 $10^{-2}$       &1.350 & 244   &  -0.0015         \\
 $10^{-4}$       &0.955 & 210   &  -0.0013         \\
 $10^{-6}$       &0.770 & 196   &  -0.0013         \\ \hline
\end{tabular}
\end{center}
\end{table}
\begin{table}[ht]
\caption{\label{Tablechi2} The numerical results for models 2 and 3 with Eq.~(\ref{fpi1}) used to
calculate $f_\pi$.}
\begin{center}
\begin{tabular}{ccccccc}
\hline\multicolumn{3}{c}{model 2}& & \multicolumn{3}{c}{model 3} \\ \cline{1-3} \cline{5-7}
$\Delta\:[{\rm GeV}^2]$ & $\lambda\:[{\rm GeV}]$ & $\Pi_{\chi}(0)/12\:[{\rm GeV}^2]$ & &
$\Delta\:[{\rm GeV}^4]$ & $\lambda\:[{\rm GeV}]$ & $\Pi_{\chi}(0)/12\:[{\rm GeV}^2]$\\ \hline
 2.000& 2.94& -0.0021& &$10^{-1}$& 1.71& -0.0016   \\
 0.200& 1.51& -0.0015& &$10^{-4}$& 0.95& -0.0012   \\
 0.020& 1.44& -0.0012& &$10^{-7}$& 0.71& -0.0012   \\ \hline
\end{tabular}
\end{center}
\end{table}
\begin{table}[htb]
\caption{\label{Tablechi3} The numerical results for models 2 and 3 with Eq.~(\ref{fpi3}) used to
calculate $f_\pi$.}
\begin{center}
\begin{tabular}{ccccccc}
\hline\multicolumn{3}{c}{model 2}&& \multicolumn{3}{c}{model 3}\\ \cline{1-3}\cline{5-7}
$\Delta\:[{\rm GeV}^2]$ & $\lambda\:[{\rm GeV}]$ & $\Pi_{\chi}(0)/12\:[{\rm GeV}^2]$ & &
$\Delta\:[{\rm GeV}^4]$ & $\lambda\:[{\rm GeV}]$ & $\Pi_{\chi}(0)/12\:[{\rm GeV}^2]$\\
\cline{1-3}\hline
 2.000& 2.955& -0.0023& &$10^{-1}$& 1.755& -0.0018   \\
 0.200& 1.600& -0.0017& &$10^{-4}$& 1.010& -0.0014   \\
 0.020& 1.570& -0.0015& &$10^{-7}$& 0.765& -0.0014   \\ \hline
\end{tabular}
\end{center}
\end{table}

The results for model 2 and model 3 are given in Table~\ref{Tablechi2}. It is shown that the values
of the quantity $\Pi_{\chi}(0)/12$ are still very small as that of Ref.~\cite{HH03}.

To further check if the second term in the brackets of Eq.~(\ref{fpi1}) give some serious
modifications to our results, we drop this term intentionally and obtain:
\begin{equation}  \label{fpi3}
   {f_\pi}^2=\frac{3}{4\pi^2}\int_0^\infty s{\rm d}s\ \frac{A^2(s)B^2(s)}{[sA^2(s)+B^2(s)]^2}.
\end{equation}
Let $A(s)=1$, $B(s)=M$, the result of Ref.~\cite{WM98} is strictly reproduced. With
Eq.~(\ref{fpi3}), similar calculations can be performed for the various models as previously did.
For example, for models 2 and 3, the results are given in Table~\ref{Tablechi3}. One can see that,
these results do not make much difference to the previous ones.

To summarize, we have calculated the QCD vacuum tensor susceptibility based on the modified version
of calculating pion decay constant. The various calculations in this letter and in Ref.~\cite{HH03}
show that, it is a fact that the value of tensor susceptibility calculated in GCM is very small. In
these calculations, the basic characteristic ($A(s)\ge1$) of quark propagator $G_0(p)$ or quark
self-energy $\Sigma(p)$ determined from the Dyson-Schwinger equations keeps unchanged. This is
probably the main reason to the small value of tensor susceptibility. So, for the calculation of
QCD vacuum tensor susceptibility, GCM formulated to date deviates seriously from the QCD sum rules
and constituent quark model.

\bigskip
\noindent {\large \bf Acknowledgements}
\bigskip

Part of this work was finished in Department of Physics, Nanjing University. The author would like
to thank Profs. Jia-lun Ping, Hong-shi Zong and Fan Wang for sincere help and valuable discussions.

\end{document}